Comment on the paper "**Photoionization of endohedral atoms using R-matrix methods: application to Xe@C$_{60}$**" by T. W. Gorczyca, M. F. Hasoglu and S. T. Manson, Phys. Rev. A**86**, 033204 (2012) [1].

M. Ya. Amusia[1, 21] and L. V. Chernysheva[2]

[1] Racah institute of Physics, the Hebrew University, Jerusalem, 91904, Israel
[2] Ioffe Physical-Technical institutes, RAS, St. Petersburg, 194021, Russia

**Abstract:** We demonstrate serious misrepresentation of existing experimental data and important omissions in the List of references in the paper [1]. We demonstrate that what is called in [1] total photoionization cross-section is in fact a partial photoionization cross-section. We demonstrate that long before [1] and presented there experimental data for photoionization of $Xe@C_{60}^{+}$ were obtained, a theoretical prediction for $Xe@C_{60}$ have been published that was in a reasonable agreement with experiment. These data are even not mentioned in [1].

Incidentally, we run across the paper [1] that is dedicated to describing R-matrix approach to endohedrals photoionization, using as a concrete object of application the endohedral $Xe@C_{60}$. The authors of [1] started from considering the total photoionization of

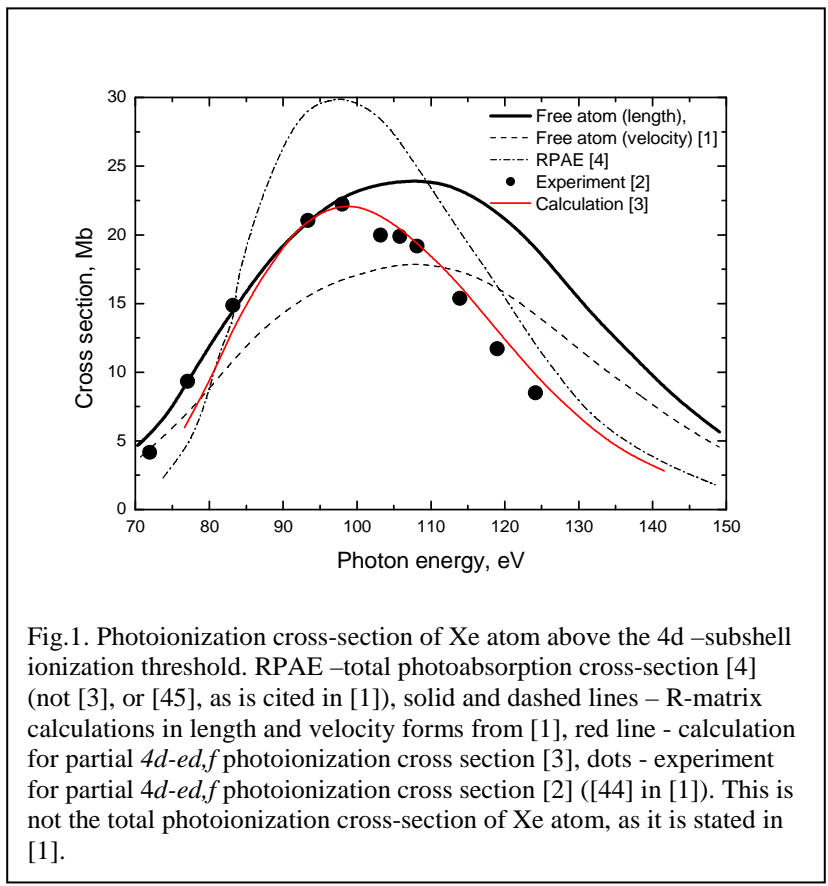

Fig.1. Photoionization cross-section of Xe atom above the 4d –subshell ionization threshold. RPAE –total photoabsorption cross-section [4] (not [3], or [45], as is cited in [1]), solid and dashed lines – R-matrix calculations in length and velocity forms from [1], red line - calculation for partial *4d-ed,f* photoionization cross section [3], dots - experiment for partial *4d-ed,f* photoionization cross section [2] ([44] in [1]). This is not the total photoionization cross-section of Xe atom, as it is stated in [1].

$Xe$ atom above its $4d^{10}$ – subshell photoionization threshold (the Fig.2 in [1]). They claim that a big discrepancy exists between experimental data from [2] (their [44]) and the RPAE calculations from [3] (their [45]). In fact, paper [1] compares RPAE results for total cross-section with partial photoionization contribution from channels $4d-\varepsilon f, p$, measured in [2]. These results, as was demonstrated in [3], should be different. The difference determines the so-called intra-atomic inelastic photoelectron scattering. The account of this effect led to demonstrated in Fig. 1 complete agreement

---

[1] E-mail: amusia@vms.huji.ac.il

between calculations [3] (solid line) and experimental data [2] (dots).

As to the total cross sections, for them very good agreement with experiment exists since long ago [4]. Our Fig. 1 is similar to Fig. 2 from [1], differing from it by an additional, solid curve from [3] (in red) that is, however, very important. This curve compared to experimental data shows that the partial photoionization cross-section of $Xe$ atom above its $4d^{10}-$ subshell was well explained in [3] more than two decades before [1] was published. To overlook this curve the authors of [1] were unable since this is the main, however omitted by them, result of cited in [1] paper [3]. Note that in [3] the reasons were clarified, why RPAE is close to the total photo-absorption cross-section and differ from contribution of only $4d-\varepsilon f, p$ channels. Namely, this is because RPAE cross-sections automatically includes two-step processes like primary 4d-photoionization as a first step with subsequent inelastic scattering of the photoelectrons $\varepsilon f, p$ upon the residual ion with 5p and 5s subshells ionization as the second step.

Now let us discuss the photoionization of $Xe@C_{60}$. The first prediction of Xe 4d Giant resonance destruction under the action of the fullerenes shell have been made in the frame of RPAE, i.e. for the total photoionization, in papers [5, 6] and confirmed at least qualitatively in [7]. There, according to the point of view of the authors of [7], the total photoionization cross-section of $Xe@C_{60}^+$ was presented. Although only one channel was in fact measured, to estimate the total cross-section the data obtained were normalized multiplying it by a factor 6.5 in order to be on the average as close as possible to the value of the experimental cross-section at the Xe 4d Giant resonance.

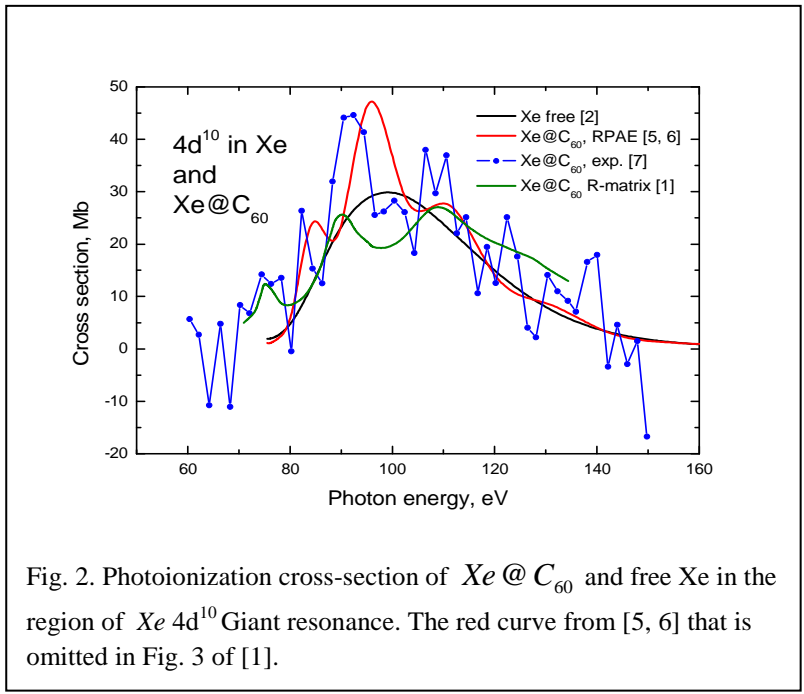

Fig. 2. Photoionization cross-section of $Xe@C_{60}$ and free Xe in the region of $Xe$ $4d^{10}$ Giant resonance. The red curve from [5, 6] that is omitted in Fig. 3 of [1].

The experiment, according to what is written in [7], was specially designed to check the validity of the predictions made in [5]. In the Letter [7] the experimental data were presented together with the results of calculations from [5, 6] along with some other calculations performed after 2005. All they were definitely further away from experiment than the results obtained in [5, 6]. The data extracted from Letter [7] are presented in [1], in their Fig.3, being however stripped of all previous calculation data, including that of [5, 6]. We are positive that it was impossible to overlook our curve from [5] while constructing Fig. 3 in [1]. Thus, we do believe that this omission in Fig. 3 of [1] was intentional. Note that the calculation result from [1] still deviates definitely not less from experiment [7] than our prediction in [5, 6]. This is illustrated in Fig.2 of this Comment. Such an omission considerably distorts the real situation by unjust enhancing the R-matrix approach achievements.

We are aware that the experiment was performed on $Xe@C_{60}^+$ instead of $Xe@C_{60}$. This makes, of course, the comparison theory – experiment more difficult. However the same problem is valid for [1] and for [5, 6], with the essential difference that [5, 6] was a prediction, while [1] followed the existing data. Even the fullerenes square well potential used in [1] was different by radiuses and depth from that generally accepted. In fact, in [1] those values of these parameters [8] were employed that permitted to reproduce data from [7].

One can say that [1] is dedicated to developing of the R-matrix approach, so it can limit itself to presenting R-matrix ability to describe existing experimental data. From our point of view, this is incorrect. A subsequent paper has to present also previous explanation of the data that are used even as a test field to demonstrate a given method's abilities. The intentional omission of the previously obtained result is a serious fault an violation of scientific ethics. Therefore, we do believe that our critical comments to the paper [1] deserved to be presented to the attention of the same readership as that of [1].